\documentclass{ws-mpla}
\usepackage{clshan-math, clshan-dm-dd}

\def \gsim {\:\raisebox{-0.7ex}{$\stackrel{\textstyle>}{\sim}$}\:}
\begin{document}
\markboth{Chung-Lin Shan}
 {Extracting Dark Matter Properties Model--Independently
  from Direct Detection Experiments}
%
\catchline{}{}{}{}{}
%
\title{Extracting Dark Matter Properties Model--Independently \\
       from Direct Detection Experiments%
}
\author{\footnotesize Chung-Lin Shan
\\~\\}
\address{Department of Physics, National Cheng Kung University \\
         No.~1, University Road,
         Tainan City 70101, Taiwan, R.O.C. \\~\\
         Physics Division, National Center for Theoretical Sciences     \\
         No.~101, Sec.~2, Kuang-Fu Road,
         Hsinchu City 30013, Taiwan, R.O.C. \\~\\
         clshan@mail.ncku.edu.tw}
%
%
%
\maketitle
\pub{Received (Day Month Year)}{Revised (Day Month Year)}
\begin{abstract}
 In this article
 I review model--independent procedures for
 extracting properties of
 Weakly Interacting Massive Particles (WIMPs)
 from direct Dark Matter detection experiments.
 Neither prior knowledge about
 the velocity distribution function of
 halo Dark Matter particles
 nor about their mass or
 cross sections on target nucleus
 is needed.
 The unique required information
 is measured recoil energies
 from experiments
 with different detector materials.
\end{abstract}
\ccode{PACS Nos.: 95.35.+d, 29.85.Fj.}
\section{Introduction}
 Different astronomical observations and measurements
 indicate that
 more than 80\% of all matter in our Universe are ``dark''
 and this Dark Matter interacts
 at most very weakly with ordinary matter.
 Weakly Interacting Massive Particles (WIMPs) $\chi$
 arising in several extensions of
 the Standard Model of particle physics
 with masses roughly between 10 GeV and a few TeV
 are one of the leading candidates for Dark Matter%
\cite{SUSYDM96,Bertone05,Steffen08,Bergstrom09}.
 Currently,
 the most promising method to detect different WIMP candidates
 is the direct detection of the recoil energy deposited
 by elastic scattering of ambient WIMPs off the target nuclei%
\cite{Smith90,Lewin96}.
 The differential event rate
 for elastic WIMP--nucleus scattering is given by%
\cite{SUSYDM96}:
\beq
   \dRdQ
 = 
   \afrac{\rho_0 \sigma_0}{2 \mchi \mrN^2}
   \FQ \int_{\vmin}^{\vmax} \bfrac{f_1(v)}{v} dv
\~.
\label{eqn:dRdQ}
\eeq
 Here $R$ is the direct detection event rate,
 i.e., the number of events
 per unit time and unit mass of detector material,
 $Q$ is the energy deposited in the detector,
 $\rho_0$ is the WIMP density near the Earth,
 $\sigma_0$ is the total cross section
 ignoring the form factor suppression,
 $F(Q)$ is the elastic nuclear form factor,
 $f_1(v)$ is the one--dimensional velocity distribution function
 of the WIMPs impinging on the detector,
 $v$ is the absolute value of the WIMP velocity
 in the laboratory frame.
 The reduced mass $\mrN$ is defined by
\beq
        \mrN
 \equiv \frac{\mchi \mN}{\mchi + \mN}
\~,
\label{eqn:mrN}
\eeq
 where $\mchi$ is the WIMP mass and
 $\mN$ that of the target nucleus.
 Finally,
 \mbox{$\vmin = \alpha \sqrt{Q}$}
 is the minimal incoming velocity of incident WIMPs
 that can deposit the energy $Q$ in the detector
 with the transformation constant
\beq
        \alpha
 \equiv \sfrac{\mN}{2 \mrN^2}
\~,
\label{eqn:alpha}
\eeq
 and $\vmax$ is the maximal WIMP velocity
 in the Earth's reference frame,
 which is related to
 the escape velocity from our Galaxy
 at the position of the Solar system,
 $\vesc~\gsim~600$ km/s.

 The total WIMP--nucleus cross section
 $\sigma_0$ in Eq.~(\ref{eqn:dRdQ})
 depends on the nature of WIMP couplings on nucleons.
 Through e.g., squark and Higgs exchanges with quarks,
 WIMPs could have a ``scalar'' interaction with nuclei. %
 The total cross section for
 the spin--independent (SI) scalar interaction
 can be expressed as\cite{SUSYDM96,Bertone05}
\beq
   \sigmaSI
 = \afrac{4}{\pi} \mrN^2 \bBig{Z f_{\rm p} + (A - Z) f_{\rm n}}^2
\~.
\label{eqn:sigma0_scalar}
\eeq
 Here $\mrN$ is the reduced mass defined in Eq.~(\ref{eqn:mrN}),
 $Z$ is the atomic number of the target nucleus,
 i.e., the number of protons,
 $A$ is the atomic mass number,
 $A-Z$ is then the number of neutrons,
 $f_{\rm (p, n)}$ are the effective
 scalar couplings of WIMPs on protons p and on neutrons n,
 respectively.
 Here we have to sum over the couplings
 on each nucleon before squaring
 because the wavelength associated with the momentum transfer
 is comparable to or larger than the size of the nucleus,
 the so--called ``coherence effect''.

 In addition,
 for the lightest supersymmetric neutralino,
 and for all WIMPs which interact primarily through Higgs exchange,
 the scalar couplings are approximately the same
 on protons and on neutrons:
\( 
        f_{\rm n}
 \simeq f_{\rm p}
.
\)
 Thus the ``pointlike'' cross section $\sigmaSI$
 in Eq.~(\ref{eqn:sigma0_scalar}) can be written as
\beq
        \sigmaSI
 \simeq \afrac{4}{\pi} \mrN^2 A^2 |f_{\rm p}|^2
 =      A^2 \afrac{\mrN}{\mrp}^2 \sigmapSI
\~,
\label{eqn:sigma0SI}
\eeq
 where $\mrp$ is the reduced mass
 of the WIMP mass $\mchi$ and the proton mass $m_{\rm p}$,
 and
\beq
   \sigmapSI
 = \afrac{4}{\pi} \mrp^2 |f_{\rm p}|^2
\label{eqn:sigmapSI}
\eeq
 is the SI WIMP--nucleon cross section.
 Here the tiny mass difference between a proton and a neutron
 has been neglected.

 On the other hand,
 through e.g., squark and Z boson exchanges with quarks,
 WIMPs could also couple to the spin of target nuclei,
 an ``axial--vector'' interaction.
 The spin--dependent (SD) WIMP--nucleus cross section
 can be expressed as\cite{SUSYDM96,Bertone05}:
\beq
   \sigmaSD
 = \afrac{32}{\pi} G_F^2 \~ \mrN^2
   \afrac{J + 1}{J} \bBig{\Srmp \armp + \Srmn \armn}^2
\~.
\label{eqn:sigma0SD}
\eeq
 Here $G_F$ is the Fermi constant,
 $J$ is the total spin of the target nucleus,
 $\expv{S_{\rm (p, n)}}$ are the expectation values of
 the proton and neutron group spins,
 and $a_{\rm (p, n)}$ are the effective SD WIMP couplings
 on protons and on neutrons.
%
 Some relevant spin values of
 the most used spin--sensitive nuclei
 are given in Table 1.

 For the SD WIMP--nucleus interaction,
 it is usually assumed that
 only unpaired nucleons contribute significantly
 to the total cross section,
 as the spins of the nucleons in a nucleus
 are systematically anti--aligned%
\footnote{
 However,
 more detailed nuclear spin structure calculations show that
 the even group of nucleons has sometimes
 also a non--negligible spin
 (see Table 1 and
  e.g., data given
  in Refs.~\refcite{SUSYDM96,Tovey00,Giuliani05,Girard05}).
}.
 Under this ``odd--group'' assumption,
 the SD WIMP--nucleus cross section can be reduced to
\beq
   \sigmaSD
 = \afrac{32}{\pi} G_F^2 \~ \mrN^2
   \afrac{J + 1}{J} \expv{S_{\rm (p, n)}}^2 |a_{\rm (p, n)}|^2
\~.
\label{eqn:sigma0SD_odd}
\eeq
 And
 the SD WIMP cross section on protons or on neutrons
 can be given as
\beq
   \sigma_{\chi {\rm (p, n)}}^{\rm SD}
 = \afrac{24}{\pi} G_F^2 \~ m_{\rm r, (p, n)}^2 |a_{\rm (p, n)}|^2
\~.
\label{eqn:sigmap/nSD}
\eeq
\begin{table}[t!]
\tbl{
 List of the relevant spin values of
 the most used spin--sensitive nuclei.
 More details can be found in
 e.g., Refs.~1, 7, 8, 9.
}{
\begin{tabular}{|| c   c   c   c   c   c   c   c ||}
\hline
\hline
 \makebox[1  cm][c]{Isotope}        &
 \makebox[0.5cm][c]{$Z$}            & \makebox[0.5cm][c]{$J$}     &
 \makebox[1  cm][c]{$\Srmp$}        & \makebox[1  cm][c]{$\Srmn$} &
 \makebox[1.2cm][c]{$-\Srmp/\Srmn$} & \makebox[1.2cm][c]{$\Srmn/\Srmp$} &
 \makebox[3  cm][c]{Natural abundance (\%)} \\
\hline
\hline
 $\rmXA{F}{19}$   &  9 & 1/2 &                   0.441  & \hspace{-1.8ex}$-$0.109 &
      4.05  &  $-$0.25   &       100   \\
\hline
 $\rmXA{Na}{23}$  & 11 & 3/2 &                   0.248  &                   0.020 &
  $-$12.40  &     0.08   &       100   \\
\hline
 $\rmXA{Cl}{35}$  & 17 & 3/2 & \hspace{-1.8ex}$-$0.059  & \hspace{-1.8ex}$-$0.011 &
   $-$5.36  &     0.19   &        76   \\
\hline
 $\rmXA{Cl}{37}$  & 17 & 3/2 & \hspace{-1.8ex}$-$0.058  &                   0.050 &
      1.16  &  $-$0.86   &        24   \\
\hline
 $\rmXA{Ge}{73}$  & 32 & 9/2 &                   0.030  &                   0.378 &
   $-$0.08  &    12.6    &  7.8 / 86 (HDMS)\cite{Bednyakov08a}\\
\hline
 $\rmXA{I}{127}$  & 53 & 5/2 &                   0.309  &                   0.075 &
   $-$4.12  &     0.24   &       100   \\
\hline
 $\rmXA{Xe}{129}$ & 54 &  1/2 &                  0.028  &                   0.359 &
   $-$0.08  &    12.8    &        26   \\
\hline
 $\rmXA{Xe}{131}$ & 54 &  3/2 & \hspace{-1.8ex}$-$0.009 & \hspace{-1.8ex}$-$0.227 &
   $-$0.04  &    25.2    &         21   \\
\hline
\hline
\end{tabular}}
\end{table}

 Due to the coherence effect with the entire nucleus
 shown in Eq.~(\ref{eqn:sigma0SI}),
 the cross section for scalar interaction
 scales approximately as the square of
 the atomic mass of the target nucleus.
 Hence,
 in most supersymmetric models,
 the SI cross section for nuclei with $A~\gsim~30$ dominates
 over the SD one\cite{SUSYDM96,Bertone05}.
\section{Reconstructing the one--dimensional
         velocity distribution function\hspace*{-0.3cm} \\ of halo WIMPs}
 As the first step of the development of
 these model--independent data analysis procedures,
 starting with a time--averaged recoil spectrum $dR / dQ$
 and assuming that no directional information exists,
 the normalized one--dimensional
 velocity distribution function of incident WIMPs, $f_1(v)$,
 is solved from Eq.~(\ref{eqn:dRdQ}) directly as\cite{DMDDf1v}
\beq
   f_1(v)
 = \calN
   \cbrac{ -2 Q \cdot \dd{Q} \bbrac{ \frac{1}{\FQ} \aDd{R}{Q} } }\Qva
\~,
\label{eqn:f1v_dRdQ}
\eeq 
 where the normalization constant $\calN$ is given by
\beq
   \calN
 = \frac{2}{\alpha}
   \cbrac{\intz \frac{1}{\sqrt{Q}}
                \bbrac{ \frac{1}{\FQ} \aDd{R}{Q} } dQ}^{-1}
\~.
\label{eqn:calN_int}
\eeq
 Note that
 the WIMP velocity distribution
 reconstructed by Eq.~(\ref{eqn:f1v_dRdQ})
 is {\em independent} of the local WIMP density $\rho_0$
 as well as of the WIMP--nucleus cross section $\sigma_0$.

 However,
 in order to use the expressions
 (\ref{eqn:f1v_dRdQ}) and (\ref{eqn:calN_int})
 for reconstructing $f_1(v)$,
 one needs a functional form for the recoil spectrum $dR / dQ$.
 In practice
 this requires usually a fit to experimental data
 and data fitting will re--introduce some model dependence
 and make the error analysis more complicated.
 Hence,
 expressions that allow to reconstruct $f_1(v)$
 directly from experimental data
 (i.e., measured recoil energies)
 have been developed%
\cite{DMDDf1v}.
 Considering experimental data described by
\beq
     {\T Q_n - \frac{b_n}{2}}
 \le \Qni
 \le {\T Q_n + \frac{b_n}{2}}
\~,
     ~~~~~~~~~~~~ 
     i
 =   1,~2,~\cdots,~N_n,~
     n
 =   1,~2,~\cdots,~B.
\label{eqn:Qni}
\eeq
 Here the total energy range between $\Qmin$ and $\Qmax$
 has been divided into $B$ bins
 with central points $Q_n$ and widths $b_n$.
 In each bin,
 $N_n$ events will be recorded.
 Since the recoil spectrum $dR / dQ$ is expected
 to be approximately exponential,
 in order to approximate the spectrum
 in a rather wider range,
 the following {\em exponential} ansatz
 for the {\em measured} recoil spectrum
 ({\em before} normalized by the exposure $\calE$)
 in the $n$th $Q-$bin has been introduced\cite{DMDDf1v}:
\beq
        \adRdQ_{{\rm expt}, \~ n}
 \equiv \adRdQ_{{\rm expt}, \~ Q \simeq Q_n}
 \equiv \rn  \~ e^{k_n (Q - Q_{s, n})}
\~.
\label{eqn:dRdQn}
\eeq
 Here $r_n = N_n / b_n$ is the standard estimator
 for $(dR / dQ)_{\rm expt}$ at $Q = Q_n$,
 $k_n$ is the logarithmic slope of
 the recoil spectrum in the $n$th $Q-$bin,
 which can be computed numerically
 from the average value of the measured recoil energies
 in this bin:
\beq
   \bQn
 \equiv \frac{1}{N_n} \sumiNn \abrac{\Qni - Q_n}
 = \afrac{b_n}{2} \coth\afrac{k_n b_n}{2}-\frac{1}{k_n}
\~.
\label{eqn:bQn}
\eeq
 Then the shifted point $Q_{s, n}$
 in the ansatz (\ref{eqn:dRdQn}),
 at which the leading systematic error
 due to the ansatz
 is minimal\cite{DMDDf1v},
 can be estimated by
\beq
   Q_{s, n}
 = Q_n + \frac{1}{k_n} \ln\bfrac{\sinh(k_n b_n/2)}{k_n b_n/2}
\~.
\label{eqn:Qsn}
\eeq
 Note that $Q_{s, n}$ differs from the central point of the $n$th bin, $Q_n$.

 Now,
 substituting the ansatz (\ref{eqn:dRdQn})
 into Eq.~(\ref{eqn:f1v_dRdQ})
 and then letting $Q = Q_{s, n}$,
 we can obtain that\cite{DMDDf1v}
\beq
   f_{1, {\rm rec}}\abrac{v_{s, n} = \alpha \sqrt{Q_{s, n}}}
 = \calN
   \bBigg{\frac{2 Q_{s, n} r_n}{F^2(Q_{s, n})}}
   \bbrac{\dd{Q} \ln \FQ \bigg|_{Q = Q_{s, n}} - k_n}
\~.
\label{eqn:f1v_Qsn}
\eeq
 Here the normalization constant $\calN$
 given in Eq.~(\ref{eqn:calN_int})
 can be estimated directly from the data:
\beq
   \calN
 = \frac{2}{\alpha}
   \bbrac{\sum_{a} \frac{1}{\sqrt{Q_a} \~ F^2(Q_a)}}^{-1}
\~,
\label{eqn:calN_sum}
\eeq
 where the sum runs over all events in the sample.
\section{Determining the WIMP mass and
         the SI WIMP--nucleon coupling\hspace*{-0.35cm}}
 By using expressions (\ref{eqn:f1v_dRdQ})
 and (\ref{eqn:calN_int})
 for reconstructing the WIMP velocity distribution function,
 not only the overall normalization constant $\calN$
 given in Eq.~(\ref{eqn:calN_int}),
 but also the shape of the velocity distribution,
 through the transformation $Q = v^2 / \alpha^2$
 in Eq.~(\ref{eqn:f1v_dRdQ}),
 depends on the WIMP mass $\mchi$
 involved in the coefficient $\alpha$.
 It is thus crucial to develop a method
 for determining the WIMP mass model--independently.

 From Eq.~(\ref{eqn:f1v_dRdQ})
 and using the exponential ansatz in Eq.~(\ref{eqn:dRdQn}),
 the moments of the normalized one--dimensional
 WIMP velocity distribution function
 can be estimated by\cite{DMDDmchi}
\beqn
    \expv{v^n}
 &=& \int_{v(\Qmin)}^{v(\Qmax)} v^n f_1(v) \~ dv
    \non\\
 &=& \alpha^n
    \bfrac{2 \Qmin^{(n+1)/2} r(\Qmin) / \FQmin + (n+1) I_n(\Qmin, \Qmax)}
          {2 \Qmin^{   1 /2} r(\Qmin) / \FQmin +       I_0(\Qmin, \Qmax)}
\~.
\label{eqn:moments}
\eeqn
 Here $v(Q) = \alpha \sqrt{Q}$,
 $Q_{\rm (min, max)}$ are the experimental
 minimal and maximal cut--off energies,
\beq
        r(\Qmin)
 \equiv \adRdQ_{{\rm expt},\~Q = \Qmin}
 =      r_1  \~ e^{k_1 (\Qmin - Q_{s, 1})}
\label{eqn:rmin}
\eeq
 is an estimated value of the {\em measured} recoil spectrum
 $(dR / dQ)_{\rm expt}$ ({\em before}
 the normalization by the exposure $\cal E$) at $Q = \Qmin$,
 and $I_n(\Qmin, \Qmax)$ can be estimated through the sum:
\beq
   I_n(\Qmin, \Qmax)
 = \sum_a \frac{Q_a^{(n-1)/2}}{F^2(Q_a)}
\~,
\label{eqn:In_sum}
\eeq
 where the sum runs again over all events in the data set.
 Note that
 by using Eq.~(\ref{eqn:moments})
 $\expv{v^n}$ can be determined
 independently of the local WIMP density $\rho_0$,
 of the WIMP--nucleus cross section $\sigma_0$,
 as well as of the velocity distribution function
 of incident WIMPs, $f_1(v)$.

 By requiring that
 the values of a given moment of $f_1(v)$
 estimated by Eq.~(\ref{eqn:moments})
 from two experiments
 with different target nuclei, $X$ and $Y$, agree,
 $\mchi$ appearing in the prefactor $\alpha^n$
 on the right--hand side of Eq.~(\ref{eqn:moments})
 can be solved as%
\cite{DMDDmchi-SUSY07}:%
\beq
   \left. \mchi \right|_{\Expv{v^n}}
 = \frac{\sqrt{\mX \mY} - \mX (\calR_{n, X} / \calR_{n, Y})}
        {\calR_{n, X} / \calR_{n, Y} - \sqrt{\mX / \mY}}
\~,
\label{eqn:mchi_Rn}
\eeq
 where
\beqn
        \calR_{n, X}
 \equiv \bfrac{2 \QminX^{(n+1)/2} r_X(\QminX) / \FQminX + (n+1) \InX}
              {2 \QminX^{   1 /2} r_X(\QminX) / \FQminX +       \IzX}^{1/n}
\~,
\label{eqn:RnX_min}
\eeqn
 and $\calR_{n, Y}$ can be defined analogously%
\footnote{
 Hereafter,
 without special remark
 all notations defined for the target $X$
 can be defined analogously for the target $Y$
 and eventually for the target $Z$.
}.
 Here $n \ne 0$,
 $m_{(X, Y)}$ and $F_{(X, Y)}(Q)$
 are the masses and the form factors of the nucleus $X$ and $Y$,
 respectively,
 and $r_{(X, Y)}(Q_{{\rm min}, (X, Y)})$
 refer to the counting rates for detectors $X$ and $Y$
 at the respective lowest recoil energies included in the analysis.
 Note that
 the general expression (\ref{eqn:mchi_Rn}) can be used
 either for spin--independent or for spin--dependent scattering,
 one only needs to choose different form factors
 under different assumptions.

 On the other hand,
 by using the theoretical prediction that
 the SI WIMP--nucleus cross section
 dominates,
 and the fact that
 the integral over the one--dimensional WIMP velocity distribution
 on the right--hand side of Eq.~(\ref{eqn:dRdQ})
 is the minus--first moment of this distribution,
 which can be estimated by Eq.~(\ref{eqn:moments}) with $n = -1$,
 one can easily find that\cite{DMDDmchi}
\beq
   \rho_0 |f_{\rm p}|^2
 = \frac{\pi}{4 \sqrt{2}} \afrac{1}{\calE A^2 \sqrt{\mN}}
   \bbrac{\frac{2 \Qmin^{1/2} r(\Qmin)}{\FQmin} + I_0}
   \abrac{\mchi + \mN}
\~.
\label{eqn:rho0_fp2}
\eeq
 Note that
 the exposure of the experiment, $\calE$,
 appears in the denominator.
 Since the unknown factor $\rho_0 |f_{\rm p}|^2$
 on the left--hand side above
 is identical for different targets,
 it leads to a second expression for determining $\mchi$:%
\cite{DMDDmchi}
\beq
   \left. \mchi \right|_\sigma
 = \frac{\abrac{\mX / \mY}^{5/2} \mY - \mX (\calR_{\sigma, X} / \calR_{\sigma, Y})}
        {\calR_{\sigma, X} / \calR_{\sigma, Y} - \abrac{\mX / \mY}^{5/2}}
\~.
\label{eqn:mchi_Rsigma}
\eeq
 Here $m_{(X, Y)} \propto A_{(X, Y)}$ has been assumed and
\beq
        \calR_{\sigma, X}
 \equiv \frac{1}{\calE_X}
        \bbrac{\frac{2 \QminX^{1/2} r_X(\QminX)}{\FQminX} + \IzX}
\~.
\label{eqn:RsigmaX_min}
\eeq

 Remind that
 the basic requirement of the expressions for determining $\mchi$
 given in Eqs.~(\ref{eqn:mchi_Rn}) and (\ref{eqn:mchi_Rsigma}) is that,
 from two experiments with different target nuclei,
 the values of a given moment of the WIMP velocity distribution
 estimated by Eq.~(\ref{eqn:moments}) should agree.
 This means that
 the upper cuts on $f_1(v)$ in two data sets
 should be (approximately) equal%
\footnote{
 Here the threshold energies of two experiments
 have been assumed to be negligibly small.
}.
 Since $v_{\rm cut} = \alpha \sqrt{Q_{\rm max}}$,
 it requires that\cite{DMDDmchi}
\beq
   Q_{{\rm max}, Y}
 = \afrac{\alpha_X}{\alpha_Y}^2 Q_{{\rm max}, X}
\~.
\label{eqn:match}  
\eeq
 Note that
 $\alpha$ defined in Eq.~(\ref{eqn:alpha})
 is a function of the true WIMP mass.
 Thus this relation for matching optimal cut--off energies
 can be used only if $\mchi$ is already known.
 One possibility to overcome this problem is
 to fix the cut--off energy of the experiment with the heavier target,
 determine the WIMP mass by either Eq.~(\ref{eqn:mchi_Rn})
 or Eq.~(\ref{eqn:mchi_Rsigma}),
 and then estimate the cut--off energy for the lighter nucleus
 by Eq.~(\ref{eqn:match}) algorithmically\cite{DMDDmchi}.

 Furthermore,
 by combining two or three data sets
 with different target nuclei
 and making an assumption for
 the local WIMP density $\rho_0$,
 we can use Eq.~(\ref{eqn:rho0_fp2})
 to estimate the {\em squared}
 SI WIMP coupling on protons (nucleons),
 $|f_{\rm p}|^2$.%
\cite{DMDDfp2-IDM2008,DMDDfp2}
 It is important to note that
 $|f_{\rm p}|^2$ and $\mchi$
 can be estimated {\em separately} and
 from experimental data directly
 with {\em neither} prior knowledge about each other
 {\em nor} about the WIMP velocity distribution.
\section{Determining ratios between
         different WIMP--nucleon cross sections\hspace*{-0.55cm}}
\subsection{Determining the ratio between two SD WIMP couplings}
 Assuming that
 the SD WIMP--nucleus interaction dominates and
 substituting the expression (\ref{eqn:sigma0SD})
 for $\sigmaSD$ into Eq.~(\ref{eqn:dRdQ})
 for two target nuclei $X$ and $Y$,
 the ratio between two SD WIMP--nucleon couplings
 can be solved analytically as%
\cite{DMDDranap-DM08,DMDDidentification-DARK2009,DMDDranap}%
\beq
   \afrac{\armn}{\armp}_{\pm, n}^{\rm SD}
 =-\frac{\SpX \pm \SpY \abrac{\calR_{J, n, X} / \calR_{J, n, Y}} }
        {\SnX \pm \SnY \abrac{\calR_{J, n, X} / \calR_{J, n, Y}} }
\~,
\label{eqn:ranapSD}
\eeq
 for $n \ne 0$.
 Here I have defined
\beq
        \calR_{J, n, X}
 \equiv \bbrac{\Afrac{J_X}{J_X + 1}
               \frac{\calR_{\sigma, X}}{\calR_{n, X}}}^{1/2}
\~,
\label{eqn:RJnX}
\eeq
 with $\calR_{n, X}$ and $\calR_{\sigma, X}$
 defined in Eqs.~(\ref{eqn:RnX_min}) and (\ref{eqn:RsigmaX_min}).

 Note that,
 firstly,
 the expression (\ref{eqn:ranapSD}) for $\armn / \armp$
 is {\em independent} of the WIMP mass $\mchi$
 and the ratio can thus be determined
 from experimental data directly
 {\em without} knowing the WIMP mass.
 Secondly,
 because the couplings in Eq.~(\ref{eqn:sigma0SD}) are squared,
 we have two solutions for $\armn / \armp$ here;
 if exact ``theory'' values for ${\cal R}_{J, n , (X, Y)}$ are taken,
 these solutions coincide for
\beq
   \afrac{\armn}{\armp}_{+, n}^{\rm SD}
 = \afrac{\armn}{\armp}_{-, n}^{\rm SD}
 = \cleft{\renewcommand{\arraystretch}{1}
          \begin{array}{l l l}
           \D -\frac{\SpX}{\SnX}         \~, & ~~~~~~~~ &
           {\rm for}~\calR_{J, n, X} = 0 \~, \\ \\ 
           \D -\frac{\SpY}{\SnY}         \~, &          &
           {\rm for}~\calR_{J, n, Y} = 0 \~,
          \end{array}}
\label{eqn:ranapSD_coin}
\eeq
 which depend only on the properties of target nuclei
 (see Table 1).
 Moreover,
 it can be found from Eq.~(\ref{eqn:ranapSD}) that
 one of these two solutions has a pole
 at the middle of two coincident values,
 which depends simply on the signs of $\SnX$ and $\SnY$:
 since $\calR_{J, n, X}$ and $\calR_{J, n, Y}$ are always positive,
 if both of $\SnX$ and $\SnY$ are positive or negative,
 the ``$-$'' solution $(\armn / \armp)^{\rm SD}_{-, n}$
 will diverge and
 the ``$+$'' solution $(\armn / \armp)^{\rm SD}_{+, n}$
 will be the ``inner'' solution;
 in contrast,
 if the signs of $\SnX$ and $\SnY$ are opposite,
 the ``$-$'' solution $(\armn / \armp)^{\rm SD}_{-, n}$
 will be the ``inner'' solution.
\subsection{Determining the ratio between
            two WIMP--proton cross sections\hspace*{-0.3cm}}
 Considering a general combination of
 both the SI and SD cross sections 
 given in Eqs.~(\ref{eqn:sigma0SI}) and (\ref{eqn:sigma0SD}),
 we can find that%
\cite{DMDDranap-DM08,DMDDranap}
\beq
   \frac{\sigmaSD}{\sigmaSI}
 = \afrac{32}{\pi} G_F^2 \~ \mrp^2 \Afrac{J + 1}{J}
   \bfrac{\Srmp + \Srmn (\armn / \armp)}{A}^2 \frac{|\armp|^2}{\sigmapSI}
 = \calCp \afrac{\sigmapSD}{\sigmapSI}
\~,
\label{eqn:rsigmaSDSI}
\eeq
 where $\sigmapSD$ given
 in Eq.~(\ref{eqn:sigmap/nSD}) has been used and
\beq
        \calCp
 \equiv \frac{4}{3} \afrac{J + 1}{J}
        \bfrac{\Srmp + \Srmn (\armn/\armp)}{A}^2
\~.
\label{eqn:Cp}
\eeq
 Then
 the expression (\ref{eqn:dRdQ})
 for the differential event rate
 should be modified to
\beqn
        \adRdQ_{{\rm expt}, \~ Q = \Qmin}
 &=&     \calE
        A^2 \! \afrac{\rho_0 \sigmapSI}{2 \mchi \mrp^2} \!\!
        \bbrac{F_{\rm SI}^2(\Qmin) + \afrac{\sigmapSD}{\sigmapSI} \calCp F_{\rm SD}^2(\Qmin)}
        \non\\
 &~&    ~~~~~~~~~~~~ \times 
        \int_{v(\Qmin)}^{v(\Qmax)} \bfrac{f_1(v)}{v} dv
\~,
\label{eqn:dRdQ_SISD}
\eeqn
 where I have used Eq.~(\ref{eqn:sigma0SI}) again.
 Now by combining two targets $X$ and $Y$
 and assuming that
 the integral over the WIMP velocity distribution function
 in Eq.~(\ref{eqn:dRdQ_SISD})
 estimated by Eq.~(\ref{eqn:moments}) for each target
 with suitable experimental maximal and minimal cut--off energies
 should be (approximately) equal,
 the ratio of the SD WIMP--proton cross section
 to the SI one can be solved
 analytically as%
\cite{DMDDranap-DM08,DMDDidentification-DARK2009,DMDDranap}
\beq
   \frac{\sigmapSD}{\sigmapSI}
 = \frac{\FSIQminY (\calR_{m, X}/\calR_{m, Y}) - \FSIQminX}
        {\calCpX \FSDQminX - \calCpY \FSDQminY (\calR_{m, X} / \calR_{m, Y})}
\~,
\label{eqn:rsigmaSDpSI}
\eeq
 where I have assumed $m_{(X, Y)} \propto A_{(X, Y)}$ and defined
\beq
        \calR_{m, X}
 \equiv \frac{r_X(\QminX)}{\calE_X \mX^2}
\~.
\label{eqn:RmX}
\eeq
 Similarly,
 the ratio of the SD WIMP--neutron cross section
 to the SI one can be given analogously as%
\footnote{
 Here I assumed that $\sigmanSI \simeq \sigmapSI$.
}:
\beq
   \frac{\sigmanSD}{\sigmapSI}
 = \frac{\FSIQminY (\calR_{m, X}/\calR_{m, Y}) - \FSIQminX}
        {\calCnX \FSDQminX - \calCnY \FSDQminY (\calR_{m, X} / \calR_{m, Y})}
\~,
\label{eqn:rsigmaSDnSI}
\eeq
 with the definition
\beq
        \calCn
 \equiv \frac{4}{3} \Afrac{J + 1}{J}
        \bfrac{\Srmp (\armp/\armn) + \Srmn}{A}^2
\~.
\label{eqn:Cn}
\eeq
 Note here that
 one can use expressions
 (\ref{eqn:rsigmaSDpSI}) and (\ref{eqn:rsigmaSDnSI})
 {\em without} a prior knowledge of the WIMP mass $\mchi$.
 Moreover,
 $\sigma_{\chi, ({\rm p, n})}^{\rm SD} / \sigmapSI$
 are functions of only $\calR_{m, (X, Y)}$,
 or, equivalently,
 the counting rate
 at the experimental minimal cut--off energies,
 $r_{(X, Y)}(Q_{{\rm min}, (X, Y)})$,
 which can be estimated with events
 in the lowest available energy ranges.

 On the other hand,
 for the general combination of
 the SI and SD WIMP--nucleon cross sections,
 by introducing {\em a third nucleus}
 with {\em only} the SI sensitivity:
\mbox{
\(
   \Srmp_Z
 = \Srmn_Z
 = 0
,
\)}
 i.e.,
\(
   {\cal C}_{{\rm p}, Z}
 = 0
.
\)
 The $\armn / \armp$ ratio can in fact be solved analytically as%
\cite{DMDDranap-DM08,DMDDidentification-DARK2009,DMDDranap}:
\beq
    \afrac{\armn}{\armp}_{\pm}^{\rm SI + SD}
 = \frac{-\abrac{\cpX \snpX - \cpY \snpY}
          \pm \sqrt{\cpX \cpY} \vbrac{\snpX - \snpY}}
         {\cpX \snpX^2 - \cpY \snpY^2}
\~.
\label{eqn:ranapSISD}
\eeq
 Here I have defined
\cheqna
\beqn
        \cpX
 &\equiv& \frac{4}{3} \Afrac{J_X + 1}{J_X} \bfrac{\SpX}{A_X}^2
        \non\\
 &~&    ~ \times \!
        \bbrac{  \FSIQminZ \afrac{\calR_{m, Y}}{\calR_{m, Z}} \!
               - \FSIQminY} \!
        \FSDQminX
\~,
\label{eqn:cpX}
\eeqn
\cheqnb
\beqn
        \cpY
 &\equiv& \frac{4}{3} \Afrac{J_Y + 1}{J_Y} \bfrac{\SpY}{A_Y}^2
        \non\\
 &~&    ~ \times \!
        \bbrac{  \FSIQminZ \afrac{\calR_{m, X}}{\calR_{m, Z}} \!
               - \FSIQminX} \!
        \FSDQminY
\~;
\label{eqn:cpY}
\eeqn
\cheqn
 and
\(
        \snpX
 \equiv \SnX / \SpX 
.
\)
 Note that,
 firstly,
 $(\armn / \armp)_{\pm}^{\rm SI + SD}$ and $c_{{\rm p}, (X, Y)}$
 given in Eqs.~(\ref{eqn:ranapSISD}), (\ref{eqn:cpX}), and (\ref{eqn:cpY})
 are functions of only
 $r_{(X, Y, Z)}(Q_{{\rm min}, (X, Y, Z)})$,
 which can be estimated with events
 in the lowest available energy ranges.
 Secondly,
 while the decision of the inner solution of
 $(\armn / \armp)_{\pm, n}^{\rm SD}$
 depends on the signs of $\SnX$ and $\SnY$,
 the decision with $(\armn / \armp)_{\pm}^{\rm SI + SD}$
 depends {\em not only} on the signs of
 $\snpX = \SnX / \SpX$ and $\snpY = \SnY / \SpY$,
 {\em but also} on the {\em order} of the two targets.

 Moreover,
 since in the expression (\ref{eqn:rsigmaSDpSI})
 for the ratio of two WIMP--proton cross sections
 there are four sources contributing statistical uncertainties,
 i.e., ${\cal C}_{{\rm p}, (X, Y)}$ and $\calR_{m, (X, Y)}$,
 in order to reduce the statistical error,
 one can choose at first a nucleus
 with {\em only} the SI sensitivity
 as the second target:
\(
   \SpY
 = \SnY
 = 0
,
\)
 i.e.,
\(
   {\cal C}_{{\rm p}, Y}
 = 0.
\)
 Then the expression in Eq.~(\ref{eqn:rsigmaSDpSI})
 can be reduced to\cite{DMDDranap}
\beq
   \frac{\sigmapSD}{\sigmapSI}
 = \frac{\FSIQminY (\calR_{m, X} / \calR_{m, Y}) - \FSIQminX}
        {\calCpX \FSDQminX}
\~.
\label{eqn:rsigmaSDpSI_even}
\eeq
 Secondly,
 one chooses a nucleus with (much) larger
 proton (or neutron) group spin
 as the first target:
\(
        \SpX
 \gg    \SnX
 \simeq 0
.
\)
 Now $\calCpX$ given in Eq.~(\ref{eqn:Cp})
 becomes (almost) independent of $\armn/\armp$:%
\beq
        \calCpX
 \simeq \frac{4}{3} \Afrac{J_X + 1}{J_X} \bfrac{\SpX}{A_X}^2
\~.
\label{eqn:CpX_p}
\eeq
\section{Summary and conclusions}
 In this article
 I reviewed the data analysis procedures
 for extracting properties of WIMP--like Dark Matter particles
 from direct detection experiments.
 These methods are model--independent
 in the sense that 
 neither prior knowledge about
 the velocity distribution function of halo Dark Matter
 nor their mass and cross sections on target nucleus
 is needed.
 The unique required information
 is measured recoil energies
 from experiments
 with different target materials.

 Once two or more experiments
 observe a few tens recoil events
 (in each experiment),
 one could in principle already estimate
 the mass and the SI coupling on nucleons
 as well as ratios between different cross sections
 of Dark Matter particles.
 All this information
 (combined eventually results from collider
  and/or indirect detection experiments)
 could then allow us to distinguish different candidates
 for (WIMP--like) Dark Matter particles
 proposed in different theoretical models
 and to extend our understanding
 on particle physics.
\section*{Acknowledgments}
 This work
 was partially supported
 by the BK21 Frontier Physics Research Division under project
 no.~BA06A1102 of Korea Research Foundation,
 by the National Science Council of R.O.C.
 under contract no.~NSC-98-2811-M-006-044,
 as well as by
 the LHC Physics Focus Group,
 National Center of Theoretical Sciences, R.O.C..
%
%

%
\end{document}